\documentclass{article}
\usepackage{spconf,amsmath,graphicx,booktabs,verbatim}
\usepackage{xcolor}

\title{Remember the context! ASR slot error correction through Memorization}
\name{Dhanush Bekal, Ashish Shenoy, Monica Sunkara, Sravan Bodapati, Katrin Kirchhoff}
\address{Amazon AWS AI, USA \\
\tt{\{dkannang, ashenoy, sunkaral, sravanb, katrinki\}@amazon.com}}
\begin{document}
%
\maketitle
\begin{abstract}

Accurate recognition of slot values such as domain specific words or named entities by automatic speech recognition (ASR) systems forms the core of the Goal-oriented Dialogue Systems. Although it is a critical step with direct impact on downstream tasks such as language understanding, many domain agnostic ASR systems tend to perform poorly on domain specific or long tail words. They are often supplemented with slot error correcting systems but it is often hard for any neural model to directly output such rare entity words. To address this problem, we propose $k$-nearest neighbor ($k$-NN) search that outputs domain-specific entities from an explicit datastore. We improve error correction rate by conveniently augmenting a pretrained joint phoneme and text based transformer sequence to sequence model with $k$-NN search during inference. We evaluate our proposed approach on five different domains containing long tail slot entities such as full names, airports, street names, cities, states. Our best performing error correction model shows a relative improvement of 7.4\% in word error rate (WER) on rare word entities over the baseline and also achieves a relative WER improvement of 9.8\% on an out of vocabulary (OOV) test set.

\end{abstract}
\section{Introduction}
\label{sec:intro}


Over the recent years, open domain Automatic Speech Recognition (ASR) models have become very powerful tools pushing the state-of-the-art performance in dialogue systems \cite{zhang2020pushing, qiantong:20, das21b_interspeech}. However, these domain agnostic models might rely on limited vocabulary during training and recognition. When such a model is used for recognizing domain specific long tail and rare word entities such as street names, retail business names, email domain names, first names, last names etc., the outputs gets mapped to the most similar sounding words in the recognition lexicon (pronunciation dictionaries). When these ASR models are used as front ends in end to end goal oriented dialogue systems, failure to recognize slots / entities leads to failure in dialogue state change.


Various methods have been proposed to mitigate this issue, which include using mixture of domain experts \cite{irie:18},  context based interpolation weights \cite{raju:18} and second-pass rescoring through domain-adapted models \cite{linda:21} to feature based domain adaptation \cite{Hentschel2019FeatureBD}. In \cite{shenoy:21, Shenoy_2021}, user-provided speech patterns were leveraged for on-the-fly adaptation. Yet another way to solve this problem is called ASR error correction. This task is modelled as machine translation problem, where one maps incorrect ASR outputs to domain words. Since ASR outputs are readily available together with the ground truth as parallel data, one can directly train sequence-to-sequence (seq2seq) models \cite{mani2020asr, mani2020towards, hrinchuk2020correction} and use them as post transcription editors on top of the domain-general ASR outputs.


In machine translation, \cite{khandelwal2020nearest} showed that contexts which are close in representation space are more likely to be followed by same target word. Further, in \cite{khandelwal2020generalization}, it was shown that learning similarity between sequence of texts is easier for a language model than predicting the next word (particularly the one from the long tail). On a similar note, we identify that seq2seq models in ASR error correction tasks find it difficult to correct ASR output to a very rare domain-specific word. We hypothesize that it might be easier for the model to explicitly memorize the speech and text pattern in an external datastore and then retrieve it using k-nearest neighbours approach to boost the softmax scores of such rare words. Following this idea, we introduce $k$-PAT (k-nearest neighbours based Phone Augmented Transformer) and demonstrate improvements in word error rate and slot accuracy across four different domains. We summarize the major contributions of our work as follows
\begin{itemize}
    \item We propose a novel method to adapt $k$-NN search for error correction task by memorizing both phone and text based representations that can be used on top of any error correcting model without the need of any additional tuning or training.
    \item We demonstrate the benefits of using such a datastore with significant improvements in WER and accuracy, particularly for rare words and out of vocabulary entities, which are critical for downstream goal-oriented dialog system tasks.
    \item We explore the effects of using lightweight domain specific datastores for efficient and fast adaptation of ASR error correction models on-the-fly.
\end{itemize}


\label{sec:models}

\begin{figure*}[th!]
  \centering
  \includegraphics[scale=1.0]{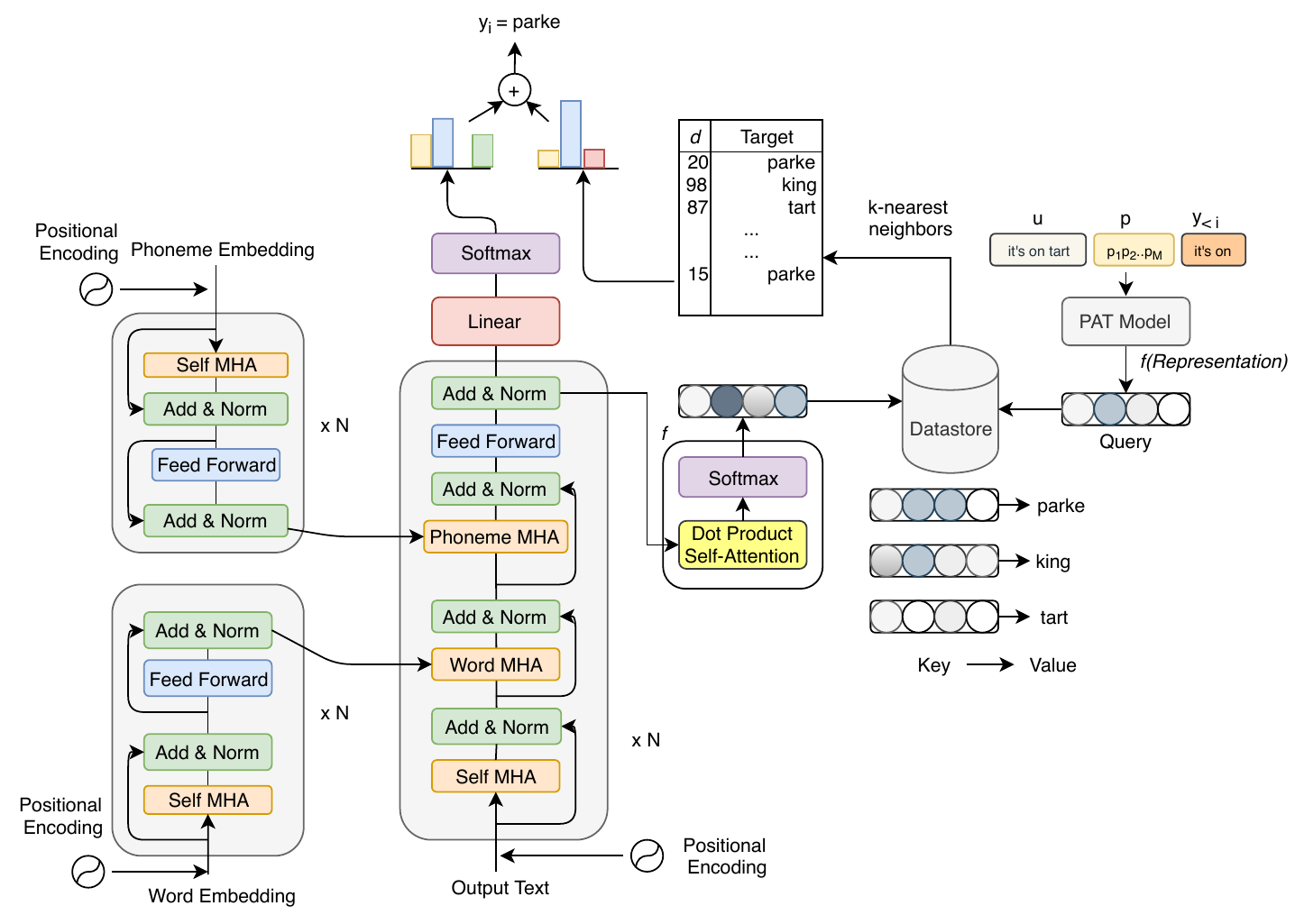}
  \caption{Proposed PAT model with $k$-NN datastore augmentation. The left half of the figure is the PAT model and the botton right half is the $k$-NN datastore creation. The top half of the model shows the inference logic. }
  \label{fig:model1}
\end{figure*}

\section{Related Work}
\label{sec:relwork}
ASR error correction task is a well studied problem in literature and usually, it has been treated as a post processing task along with other tasks like punctuation prediction \cite{sunkara2020robust, sunkara2020multimodal} and inverse text normalization \cite{sunkara2021neuralitn}. The prior works have explored the problem using a variety of subtasks including grammar error correction, improving human readability \cite{liao2021generating}, entity retrieval \cite{wang2020asr} etc. In \cite{guo2019spelling}, the authors use an RNN based external language model along with a stacked RNN based seq2seq spelling correction model for improving a baseline Listen Attend and Spell (LAS) based ASR system. Recently, pre-trained language models such as BERT, Transformer-XL and RoBERTa have been applied to the task and have been shown to be effective in making significant gains in WER  \cite{hrinchuk2020correction, liao2021generating}. The work in \cite{mani2020asr} adapts an open domain google ASR model to medical domain using a Transformer based seq2seq model. All these works also show qualitative improvements in readability, grammatical and semantic correctness of corrected sentences to support the quantitative metrics. However, all the above approaches model ASR error correction task as a text only approach. ASR also consists of phonetic information which can be used to disambiguate and gain further improvements.




In \cite{wang2020asr}, an Augmented Transformer model is proposed which leverages phonetic along with text for correcting ASR outputs. They show that jointly encoding both phoneme and text information helps in improving entity retrieval compared to a vanilla text transformer. In this work, we use this as our base model and augment $k$-NN datastore creation and inference techniques \cite{khandelwal2020nearest, khandelwal2020generalization} for improving the model without any further training. To this end, we generalize $k$-NN datastore creation idea \cite{khandelwal2020generalization} from storing just context based text patterns to using joint text and speech contextual information that provides us an additional leverage in task-oriented dialogue systems. In addition to that, we bring in a novel idea and experimentation on smaller domain specific datastores which none of the existing works cover.



\section{Models}
\subsection{Phone Augmented Transformer}
Following \cite{wang2020asr}, we implement a Phone Augmented Transformer (PAT) which consists of phonetic and text encoders and a joint decoder. The encoders and the decoder are made of stacked multi headed self attention layers \cite{vaswani2017attention}.
 The input of the transformer model is an ASR output word sequence $U = \{u_1,u_2,...u_N\}$ tokenized by sentence-piece tokenization\cite{kudo2018sentencepiece} and a phoneme sequence $P = \{p_1, p_2,...p_M\}$ generated from the ASR output words using the ASR lexicon. The output of the model is the rewritten ASR output represented by a new sequence of words $W = \{w_1, w_2,...w_O\}$.

The Phonetic encoder transforms the phone sequence $P$ into a sequence of hidden representations \\
$H_p=\{h_{p_1},h_{p_2},...h_{p_M}\}$ and the text encoder encodes the sequence of tokenized words $U$ into hidden representations $H_u=\{h_{u_1},h_{u_2},...h_{u_N}\}$

The decoder also consists of stacked multi-headed self attention layers. Each decoder self attention layer has a decoder self attention, and two encoder-decoder self attentions. The two encoder-decoder attention layers are used to fuse both input text and phonetic information. The encoder-decoder attention layers are sequentially stacked on top of the decoder attention layer. First, the text encoder-decoder attention $A_u$ between encoder output $H_u$ and the decoder attention output $H_w$ is computed using:

\begin{equation}\label{eq:txt_attn}
\begin{aligned}
A_u = softmax(\frac{(Q_u H_u)(K_u H_u)}{(T \sqrt{d_k})}) (V_u H_w)
\end{aligned}
\end{equation}

where $Q_u,K_u,V_u \epsilon R^{d_k×d_k}$ are  the  parameters  of  the attention layer,  and $d_k$ is  the  dimension  of  the  output  from  the  self-attention layer

We then fuse the phone encoder output $H_p$ with the previous attention layer output $A_u$ using:

\begin{equation}\label{eq:ph_attn}
\begin{aligned}
A_p = softmax(\frac{(Q_p H_p)(K_p H_p)}{(T \sqrt{d_k})}) (V_p A_u)
\end{aligned}
\end{equation}

where $Q_p,K_p,V_p \epsilon R^{d_k×d_k}$ are the parameters of the attention layer.

The attention output is then passed through a feed forward layer and a layer normalization layer to generate the decoder outputs. These outputs can be fed to another layer or be used as the final output representations for predicting the output word sequence. The final decoder output consists of a sequence of hidden representations $D_w=\{d_{w_1},d_{w_2},...d_{w_O}\}$ which is passed through a feed forward layer followed by a softmax layer to predict the tokenized target words $W = \{w_1, w_2,...w_O\}$.
For training the Transformer model, we follow vanilla Transformer training techniques except in our case, we also have an additional encoder. Additionally, we use the text input dropout and clean input techniques mentioned in \cite{wang2020asr}.

\subsection{Phone Augmented Transformer with k-NN Search}
In this section, we propose the $k$-nearest-neighbors ($k$-NN) retrieval augmentation to the PAT model calling it $k$-PAT. The $k$-PAT model is an inference augmentation to the PAT model based on the works of $k$-NN language model \cite{khandelwal2020generalization} and $k$-NN Machine Translation \cite{khandelwal2020nearest}. Figure \ref{fig:model1} shows a schematic overview of the proposed model. The crux of the model is in a memorization step before inference and a score interpolation step during inference.
The decoder output representations $D_w$ of all training data are memorized into an external datastore during memorization step. During inference step, the $k$ nearest neighbors of the decoder output representations are retrieved from the cached datastore. A softmax distribution computed over these retrieved neighbors is interpolated with the decoder softmax scores as show in figure \ref{fig:model1}.

Interpolating the distribution from a large parametric neural error correction model, which inherently encodes short term memory, with a non-parametric episodic memory component like $k$-NN memory cache allows for an effective way to adapt the final probability distribution towards domain-specific rare words without having to retrain the underlying model.



\subsubsection{Memorization - Datastore creation}
The datastore creation step involves generating key value pairs $(k,v)$ for all training data and storing them in an external memory.
Let $D_w = \{d_{w_1}, d_{w_2},...d_{w_{i-1}},...d_{w_O} \}$ be the sequence of decoder output representations and $W = \{w_1,w_2,...w_i,...w_O\}$ the sequence of target words. Let $f$ be a function which maps $d_{w_{i-1}}$ to a fixed vector $e_{i-1}$. Then, given the target word $w_i$, we store $(e_{i-1}, w_i)$ as the key value pair $(k_j,v_j)$. These pairs are generated for all $d_{w_i}$ in $D_w$ and for all $D_w$ generated with the training set. This datastore we create maps the contextual information (input phonetic and text) implicitly encoded in the hidden state outputs of the decoder to the target word in the sequence.

In our work we use a dot product self-attention based pooling described in equation \ref{eq:datastore} for creating the key vector $e_{i-1}$ for the decoder output $d_{i-1}$.

\begin{equation}\label{eq:datastore}
\centering
\begin{aligned}
S = softmax(D_{w} \cdot D_w^T) \\
e_{i-1} = f(d_{w_{i-1}}) = s_{i-1} \cdot D_w \\
(k_j, v_j) = (e_{i-1},w_i)\\
\end{aligned}
\end{equation}

where $S = \{s_1, s_2,...s_{O-1} \}$ and $i \epsilon (2, O)$ is a sequence of dot product attention weights. Generally speaking, any hidden layer from the decoder can be used for generating the keys. In our experiments, we found that the key vector representation obtained from equation \ref{eq:datastore} works best. This self-attention computation allows for look ahead context for each key representation $e_{i-1}$. The look ahead helps learning information about the future decoded word. Therefore, the $k$-NN datastore created from these representations has implicit information about the word to be decoded. This form of key value generation differs from the original paper \cite{khandelwal2020nearest} where the key representation only has information about the past. During inference however, we do not use the look ahead in the self attention.





\subsubsection{Inference - $k$-NN Retrieval}
At inference time, given a source ASR output : \[u=\{u_1, u_2, ...u_N\}\]  and the phonetic sequence : \[p=\{p_1, p_2, ...p_M\}\]
the PAT model produces a distribution over vocabulary as :
\[p_{PAT}(y_i | u, p, \hat{y}_{1:i-1})\] where $y_i$ is the target token at every time step and $\hat{y}$ represents the corrected output. In parallel, the model also generates the output representation $d_{w_{i-1}}$, which is  used to generate the search key : $f(d_{w_{i-1}})$. During decoding, we use this key to retrieve the $k$ nearest neighbor keys from the datastore and the corresponding values. In our experiments, we use $L2$ distance for choosing the neighbors. Once the nearest neighbors are retrieved, we use the technique described in \cite{khandelwal2020nearest} for assigning the probability scores to the retrieved values and interpolate them with the decoder softmax scores using equation \ref{interpol}. For fast retrieval of key, value pairs during inference, we use a datastore index generated using FAISS \cite{JDH17, khandelwal2020generalization, khandelwal2020nearest}.

\begin{equation}\label{interpol}
\begin{aligned}
\begin{split}
p_{kPAT}(y_i|u, p, \hat{y}_{1:i-1}) = (1- \lambda) p_{PAT}(y_{i}|u, p, \hat{y}_{1:i-1}) \\
+ \lambda p_{kNN}(y_i|u, p, \hat{y}_{1:i-1})
\end{split}
\end{aligned}
\end{equation}

We set $\lambda$ to 0.5 in our experiments to obtain the final interpolation weights.


\begin{table*}[th!]
  \centering
  \small
    \begin{tabular}{llrrrrrr}
        \toprule
         &Domain & \multicolumn{3}{c}{PAT} & \multicolumn{3}{c}{$k$-PAT} \\
              & & WER & Recall & Accuracy & WER (WERR) & Recall & Accuracy  \\
        \midrule
        1 & Combined & 10.7 & 0.70 & 82.0 & \textbf{9.9 (7.5)} & \textbf{0.71} & \textbf{82.7}  \\
        2 & Full Names & 16.5 & 0.77 & 81.3 & \textbf{15.9 (3.6)} & \textbf{0.78} & \textbf{82.1} \\
        3 & Airports & 8.6 & 0.70 & 86.8 & \textbf{7.4 (13.9)} & \textbf{0.74} & \textbf{89.0} \\
        4 & Street Names & 10.2 & 0.68 & 78.0 & \textbf{9.9 (2.9)} & \textbf{0.69} & \textbf{79.4} \\
        5 & Cities \& States & 8.5 & 0.70 & 89.7 & \textbf{8.0 (5.9)} & \textbf{0.72} & \textbf{90.3} \\
        6 & OOV  & 34.7 & 0.21 & 46.5 & \textbf{31.3 (9.8)} & \textbf{0.30} & \textbf{52.6}  \\
     \bottomrule
    \end{tabular}
    \caption{Comparison of models on different slot types. WER - Word Error Rate, Recall - Recall on slot entities, Accuracy - Slot accuracy, OOV - Out-of-vocabulary}
    \label{tab:wer_report}
\end{table*}

\begin{table}[h!]
  \centering
  \small
    \begin{tabular}{llrrr}
        \toprule
         &Datastore & WER & Recall & Accuracy \\
        \midrule
        1 & Train & 34.7 & 0.20 & 46.0 \\
        2 & Train + OOV  & 33.6 & 0.28 & 51.0 \\
        3 & OOV  & 31.3 & 0.30 & 52.6  \\

     \bottomrule
    \end{tabular}
    \caption{Performance of $k$-PAT model on OOV data set with different datastores }
    \label{tab:oov}
\end{table}

\begin{table}[h!]
\small
\centering
    \begin{tabular}{llccc}
    \toprule
    & Domain & WER & Recall & Accuracy \\
    \midrule
    1 &Full Names & 15.9 & 0.78 & 82.1 \\
    2 &Airports & 7.2 & 0.74 & 89.0 \\
    3 &Street Names & 9.5 & 0.71 & 80.3 \\
    4 &Cities \& States & 7.8 & 0.74 & 90.7 \\
    \bottomrule
\end{tabular}
\caption{WER, Recall and Accuracy of $k$-PAT Model with  slot-specific data stores}
\label{tab:domain_specific}
\end{table}

\begin{table*}[h!]
  \centering
  \small
    \begin{tabular}{llrrrr}
        \toprule
         & Good Rewrite & Ground Truth & ASR Output & PAT & $k$-PAT \\
        \midrule
        1 & Full Names & my name is janie burdick & my name is jamie burger & my name is jami barker & my name is \textbf{janie burdick}\\
        2 & Airports & i'm at prudhoe & i dont product & i'm at crowdus & i'm at \textbf{prudhoe} \\
        3 & Street Names & i stay in bedford & i stay in that for & i stay in nagpur & i stay in \textbf{bedford} \\
        4 & Cities \& States & my house is in fairhope & my house and pharaoh & my house is in fairgrove & my house is in \textbf{fairhope}\\
        5 & Long Tail & i'm at lafon & i'm at a fun & i'm at vaughan & i'm at \textbf{lafon} \\
        6 & Long Tail & i work in debrecen & i work in the back & i work in napa & i work in \textbf{debrecen} \\
        7 & OOV & i'm from lauramie & i am from lauren i & i'm from wharton & i'm from \textbf{lauramie} \\
        8 & OOV & this is anderson driscoll & this is anderson driskell & this is andy driscoll & this is \textbf{anderson driscoll} \\
        9 & OOV & book a car to zhangxiao & book a car to saying south & book a car to zanesville & book a car to \textbf{zhangxiao}\\
        \midrule
        & Bad Rewrite & & & & \\
        \midrule
        1 & Long Tail & my home is in subang & my home is in salon & my home is in salon & my home is in \textcolor{red}{solvang} \\
        2 & OOV & my name's jaylin duran & my name is jaylon joanne & my name's jaylin ceja & my name's \textcolor{red}  {jaylin durant} \\
        3 & OOV & it's on leralynn & it's on there i land & it's on neron & it's on \textcolor{red} {lerawood} \\
     \bottomrule
    \end{tabular}
    \caption{Qualitative examples from test sets generated with PAT, $k$-PAT models}
    \label{tab:qualitative}
\end{table*}

\section{Experiments}
\label{sec:expts}

\subsection{Data}
We generate synthetic data which comprises of named entities from four domain data collection \cite{builtin} namely Airports, Cities, States, Street names. For synthetic data, we augment the entities with most likely carrier phrases as prefixes. They are then passed through a Text To Speech (TTS) system followed an ASR model for simulating ASR errors. Additionally, we have vendor collected audio datasets comprising common US first and last name entities. The sampled utterances are then randomly split into train, dev and test sets with a 80:20:20 split.

We also train a sentence piece model \cite{kudo2018sentencepiece} using the training split. Both our transformer and $k$-NN datastore are sub-word based models with a vocabulary size of 32,000. In addition, we also use an internal pronunciation lexicon dictionary, same as the one used in the ASR model, to store the pronunciation of all words in the training data. This dictionary has close to 100k pronunciations.

\subsubsection{Error simulation and synthetic data augmentation}
We use TTS for synthesizing audios comprising entities from the above mentioned slots. Each of the synthetically generated text samples are passed through three randomly selected neural voices for generating the synthetic utterances. Once the utterances are generated, we pass them through an ASR model for generating the outputs. We use the n-best lists generated by the ASR model for augmenting the training data.

\begin{figure}[!htb]

\begin{minipage}[b]{1.0\linewidth}
  \centering
  \centerline{\includegraphics[scale=0.4]{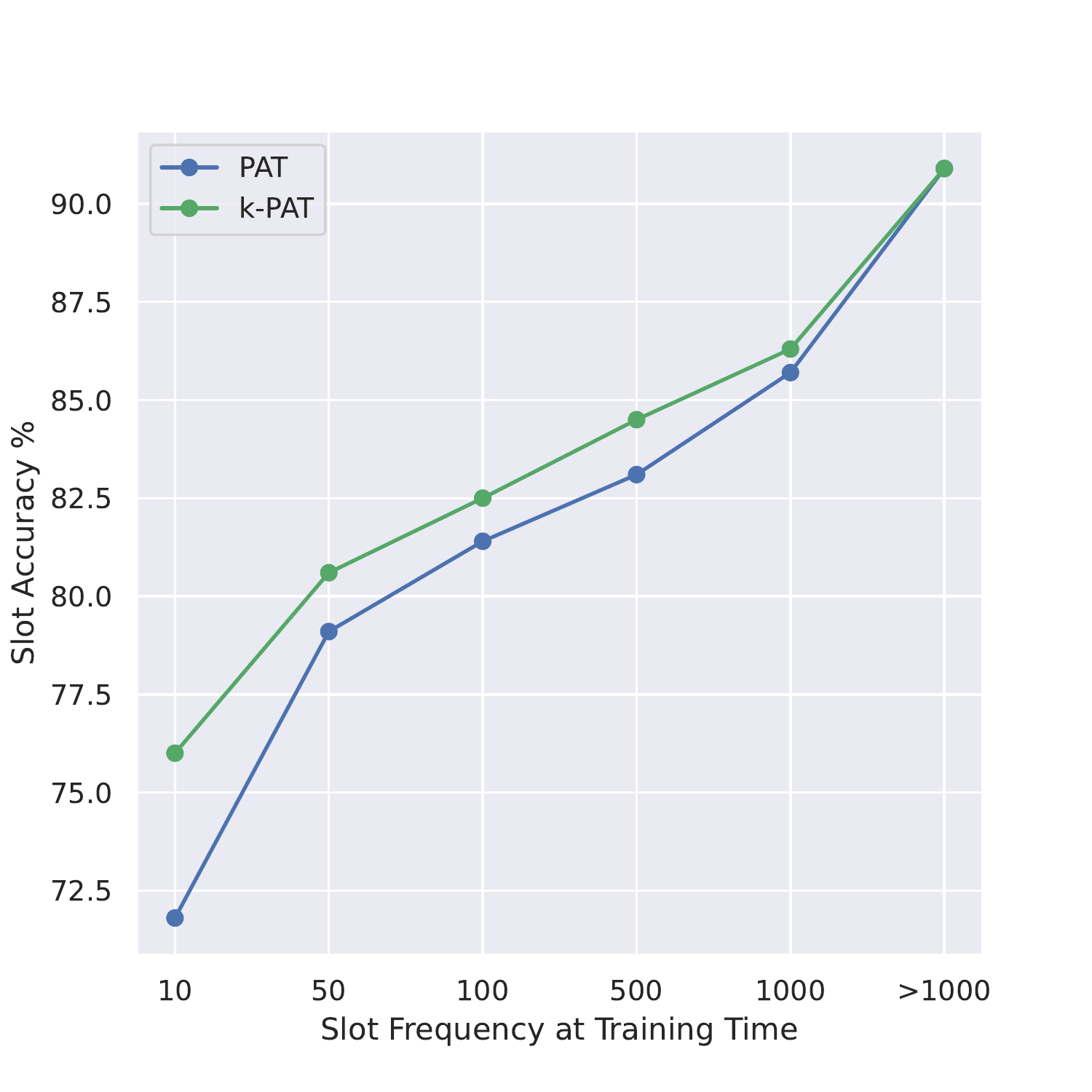}}
 \medskip
\end{minipage}
\caption{Slot accuracy vs the frequency of slots seen during training time. The biggest gain with $k$-PAT is on long-tail frequencies.}
\label{fig:slot_acc}

\begin{minipage}[b]{1.0\linewidth}
  \centering
  \centerline{\includegraphics[scale=0.4]{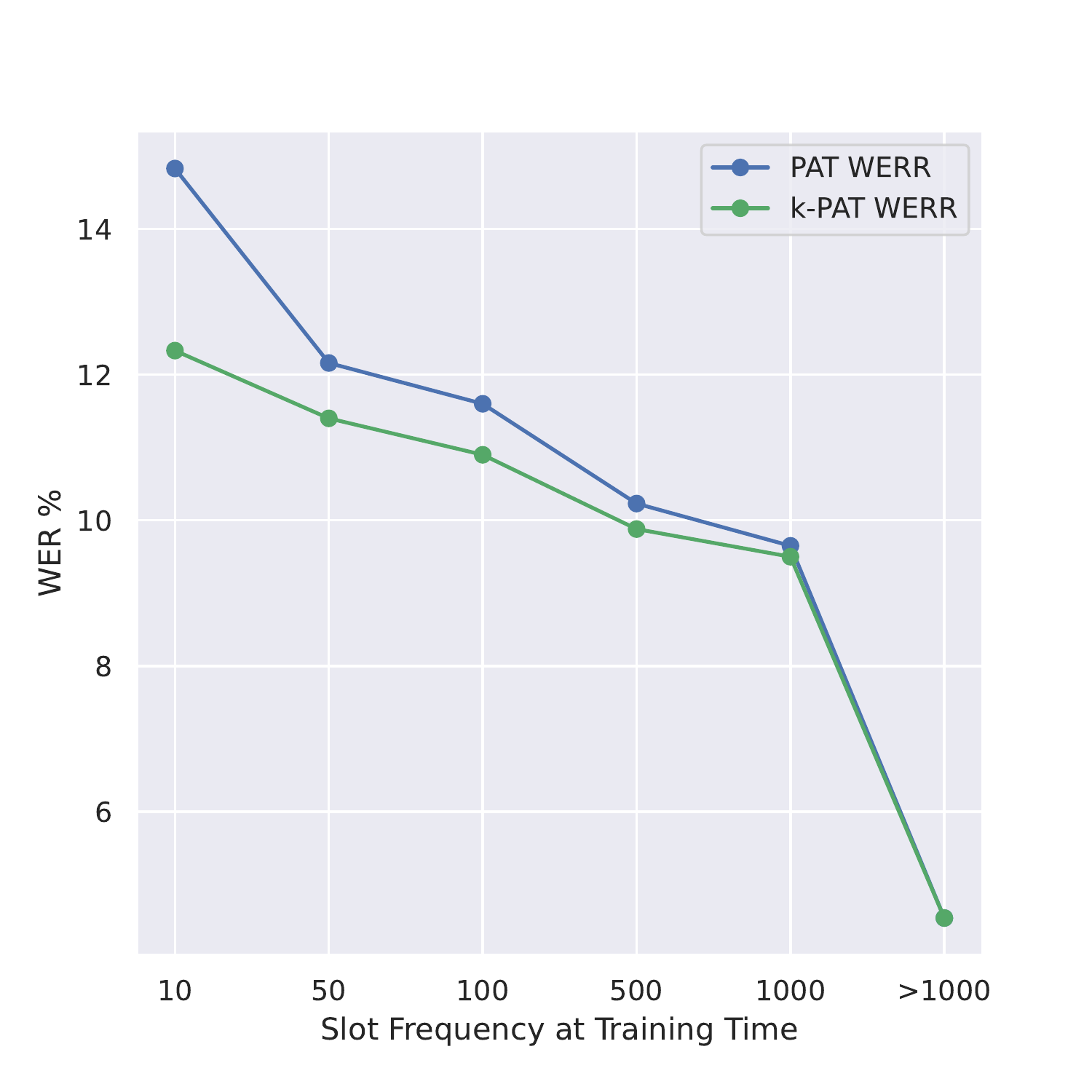}}
 \medskip
\end{minipage}
\caption{Word Error Rates (WER) with PAT and $k$-PAT for different ranges of slot frequencies seen during training time}
\label{fig:slot_wer}
\end{figure}

\subsection{Model configurations}
We use neural version of AWS Polly for generating synthetic 8kHz audios. For decoding these synthetic audios, we use a hybrid ASR system with a standard 4-gram lm based first pass decoding.

For all experiments, we use a single PAT model which consists of 4-layer encoders and a 4-layer decoder. We use 128 dimensions for the hidden representations with 8 heads for the multi headed attention in both the encoder and the decoder. The feed forward layer after the decoder has 512 dimensions. For training the model, we apply the same optimizer as the original transformer \cite{vaswani2017attention} with 4000 warm-up steps. We use a batch size of 512 and train the model for 40 epochs.  In addition to this, we follow the dropout and error free token swap-in techniques used in \cite{wang2020asr}.

The $k$-NN datastore key has 128 dimensions, same as the decoder output hidden representation size. We build the FAISS indexing \cite{JDH17} using the same techniques mentioned in \cite{khandelwal2020generalization, khandelwal2020nearest}. The number of cluster centroids for datastore creation and quantization are the same as those mentioned in \cite{khandelwal2020generalization}.
Compared to the original paper, our data store is relatively smaller with roughly 5 million key, value pairs and 5GB in memory. In addition to a combined data store, we also create slot specific data stores which have roughly 1 million key value pairs and 1GB in memory. Our experiments show that creating individual datastores leads to further improvements over a combined datastore. In all our experiments, we set k=10 for $k$-NN retrieval during inference.



\section{Results and Discussion}
\label{Results}

\textbf{Performance on different slot types}:
We compare the performance of $k$-PAT and PAT model in terms of Word  Error  Rate  (WER) and Relative Word Error Rate (WERR). We  also  calculate  recall  and  accuracy  on  the  slot  words  of  the  corrected  outputs  to  show that the $k$-PAT model is actually retrieving slot words and not just correcting the carrier phrases.  For evaluating on the slot words, we remove all the stop words (commonly used function words, such as conjunctions and preposition) from the transcriptions. Table \ref{tab:wer_report} summarizes these metrics across all our data sets of four different slot types. On the combined dataset, we observe that our $k$-PAT model achieves a 7.5\% WERR over the PAT model without any additional training. When tested on the domain specific test sets, $k$-PAT consistently outperforms PAT model across all domains in terms in WER.


Additionally, we measure slot word recognition performance by tagging slot words in the reference text for calculating recall and accuracy.
From the results shown in Table \ref{tab:wer_report}, it is clear that $k$-PAT is able to resolve slots better in the incorrect ASR output. This clearly indicates that the $k$-PAT model is effectively memorizing the context during the datastore creation step and retrieving the domain relevant neighbors during the inference step.  Qualitative examples in Table \ref{tab:qualitative} show $k$-PAT's ability to pick the correct word among similar sounding neighbors.



\noindent \textbf{Effectiveness on Long Tail Entities}: For evaluating the performance of the $k$-PAT model on long tail words, we create separate test sets from the combined test set with slot words occurring at different frequencies in the training data.
We plot the results of slot accuracy and WER against slot word frequency in figures \ref{fig:slot_acc} and \ref{fig:slot_wer} respectively. The figures indicate higher relative gains in slot accuracy and WER on long tail (low frequency) words with the $k$-PAT model. For example, at the lowest frequency of 10, we see an absolute gain of 6 percentage points in slot accuracy and an absolute reduction of 3 points in WER. This shows that $k$-PAT is effectively memorizing and correcting long tail words. Additionally, from our experiments, we observe that at higher frequencies, the gains in WER and slot-accuracy from $k$-PAT tend to be narrower and close to zero after a certain frequency threshold. This is expected as, at higher frequencies, the PAT model's softmax distribution is already biased to the high frequency word and the $k$-NN score interpolation doesn't make any additional contributions.

\noindent \textbf{Performance on OOV Data}: We evaluate the effectiveness of our proposed approach on OOV words or unseen slots during training. For this purpose, we create an out of vocabulary (OOV) dataset spanning all four domains: airports, names, streetnames, and cities, states. From Table \ref{tab:wer_report}, we see that our $k$-PAT model performs much better on OOV data with a WERR improvement of 9.8\% over the PAT model. By just memorizing unseen data, the model was able to improve on slot recovery without any additional tuning.

Additionally, we ablate on the types of datastore that can be used for OOV prediction. From Table \ref{tab:oov}, we observe that using training datastore results in negligible reduction in performance. This proves that the model doesn't retrieve any neighbors for OOV data. Additionally, we use a datastore created from OOV data alone, which gave significant improvements in both WER and slot accuracy. Furthermore, we create another datastore by combining the oov datastore with the training datastore which only gave slight improvements in WER and slot accuracy compared to the oov specific datastore. This is because the $k$-NN retrieval has more neighbors outside of OOV data to choose from which leads to higher confusion. We have shown some qualitative examples comparing PAT and $k$-PAT model in Table \ref{tab:qualitative} which show $k$-PAT retrieving the slots accurately. We also show some cases where $k$-PAT model is unable to retrieve the right neighbors. However, even in these cases, we can see that $k$-PAT model is able to retrieve retrieve similar sounding words. We intend to improve OOV performance in our future work.

\noindent \textbf{Ablation study on Slot-specific Datastores:}
We evaluate the performance of the proposed approach using individual domain specific data stores. In order to retrieve neigbors from domain specific datastores, we create individual datastores for airports, names, streetnames, and cities, states domains. Table \ref{tab:domain_specific} summarizes our results of evaluating slot specific test sets with individual data stores. Using this technique, we have observed additional performance gains in airports, street names and cities. These gains can be attributed to using a smaller $k$-NN search space and retrieval of neighbors which belong to the relevant slot being decoded. This method of storing individual datastores is very effective during run time in chat bot systems. Using the dialog state context we can retrieve domain specific data stores for correcting asr outputs in each turn.

\section{Conclusion}
In this work we have proposed an inference technique for using $k$-nearest neighbour search with a pre-trained Transformer for the task of ASR error correction. Our experiments have shown that interpolating $k$-NN scores with the decoder softmax scores provides significant gains in WER performance and slot word retrieval. We have also shown that using $k$-NN retrieval with domain specific datastores provides further gains and have an advantage over the combined datastore during runtime. This method also has the potential to dynamically augment the vocabulary of any pre-trained error correction model without retraining. In future works we will explore the impact of this method on end-to-end models and explore few shot domain adaptation.

\subsection*{Acknowledgements}
\thispagestyle{empty}
Haoyu Wang (Amazon) for providing insights into their PAT work, Saket Dingliwal (Amazon) for thoughtful discussions and providing feedback on the paper.
\clearpage



\bibliographystyle{IEEEbib}
\bibliography{strings,refs}

\end{document}